\documentclass[10pt, conference]{IEEEtran}
\IEEEoverridecommandlockouts

\usepackage{hyperref}
\usepackage[utf8]{inputenc}
\usepackage{cite}
\usepackage{amsmath,amssymb,amsfonts}
\usepackage{graphicx}
\usepackage{textcomp}
\usepackage{xcolor}
\usepackage{booktabs}
\usepackage{tikz}
\newcommand*\circled[1]{\tikz[baseline=(char.base)]{
            \node[shape=circle,draw,inner sep=2pt] (char) {#1};}}
\usepackage{soul}
\usepackage{xspace}
\usepackage{url}
\usepackage{listings,multicol,multirow}
\usepackage[caption=false]{subfig}  
\usepackage[export]{adjustbox}
\usepackage{breakurl} 
\usepackage{cleveref}
\usepackage{algorithm}
\usepackage[noend]{algpseudocode}
\usepackage{diagbox}
\usepackage{fixfoot}
\usepackage{enumerate}
\usepackage[shortlabels]{enumitem}

\newcounter{observation}

\makeatletter
\newcommand{\linebreakand}{%
  \end{@IEEEauthorhalign}
  \hfill\mbox{}\par
  \mbox{}\hfill\begin{@IEEEauthorhalign}
}
\makeatother

\usepackage{balance}

\newcommand{\ie}{\emph{i.e.,}\xspace}
\newcommand{\eg}{\emph{e.g.,}\xspace}

\newcommand{\totalsteps}{779\xspace}

\newcommand{\javasubmissions}{1,073,018\xspace}
\newcommand{\javamiddle}{428,110\xspace}
\newcommand{\javadeep}{352,859\xspace}
\newcommand{\javashallow}{292,049\xspace}

\newcommand{\javafirstattemptper}{78} 
\newcommand{\javastudents}{37,892\xspace}
\newcommand{\javasteps}{415\xspace}

\newcommand{\pythonsubmissions}{1,345,332\xspace}
\newcommand{\pythonmiddle}{618,257\xspace}
\newcommand{\pythondeep}{346,363\xspace}
\newcommand{\pythonshallow}{380,712\xspace}

\newcommand{\pythonfirstattemptper}{86} 

\newcommand{\pythonstudents}{46,863\xspace}
\newcommand{\pythonsteps}{364\xspace}

\begin{document}

\title{Analyzing the Quality of Submissions \\ in Online Programming Courses}

\author{\IEEEauthorblockN{Maria Tigina}
\IEEEauthorblockA{\textit{JetBrains Research} \\
Republic of Serbia \\
maria.tigina@jetbrains.com}

\and

\IEEEauthorblockN{Anastasiia Birillo}
\IEEEauthorblockA{\textit{JetBrains Research} \\
Republic of Serbia \\
anastasia.birillo@jetbrains.com}

\and

\IEEEauthorblockN{Yaroslav Golubev}
\IEEEauthorblockA{\textit{JetBrains Research} \\
Republic of Serbia \\
yaroslav.golubev@jetbrains.com}

\linebreakand

\IEEEauthorblockN{Hieke Keuning}
\IEEEauthorblockA{\textit{Utrecht University} \\
The Netherlands  \\
h.w.keuning@uu.nl}

\and

\IEEEauthorblockN{Nikolay Vyahhi}
\IEEEauthorblockA{\textit{Stepik} \\
United States \\
vyahhi@stepik.org}

\and

\IEEEauthorblockN{Timofey Bryksin}
\IEEEauthorblockA{\textit{JetBrains Research} \\
Republic of Cyprus \\
timofey.bryksin@jetbrains.com}
}

\maketitle

\begin{abstract}
Programming education should aim to provide students with a broad range of skills that they will later use while developing software. An important aspect in this is their ability to write code that is not only correct but also of high quality. Unfortunately, this is difficult to control in the setting of a massive open online course. In this paper, we carry out an analysis of the code quality of submissions from JetBrains Academy --- a platform for studying programming in an industry-like project-based setting with an embedded code quality assessment tool called Hyperstyle. We analyzed more than a million Java submissions and more than 1.3 million Python submissions, studied the most prevalent types of code quality issues and the dynamics of how students fix them. We provide several case studies of different issues, as well as an analysis of why certain issues remain unfixed even after several attempts. Also, we studied abnormally long sequences of submissions, in which students attempted to fix code quality issues after passing the task. Our results point the way towards the improvement of online courses, such as making sure that the task itself does not incentivize students to write code poorly.
\end{abstract}

\begin{IEEEkeywords}
programming education, code quality, MOOC, learning programming, refactoring, large-scale analysis
\end{IEEEkeywords}

\section{Introduction}\label{sec:introduction}

Nowadays, software lies at the heart of virtually every area of our life~\cite{boehm2006view}, including such crucial fields as medicine, banking, and governance, where the cost of even a single error can be incredibly high~\cite{zhong2015empirical, lions1996ariane}.
Since the quality of code directly affects the maintainability, flexibility, and performance  of the software~\cite{martin2009clean, glass2002facts, fowler2018refactoring}, maintaining its high quality also becomes a priority~\cite{fagan2002design}.
However, many novice programmers do not pay enough attention to the quality of the software they develop~\cite{keuning2017code},
which indicates the importance of instilling the ability to write high-quality code during the education process~\cite{borstler2018know, keuning2019teachers}.

A popular way of learning programming is attending massive open online courses (MOOCs)~\cite{oh2020design}, which answer the needs of a growing number of students, in particular during the pandemic~\cite{impey2021moocs}.
While MOOCs allow everyone to get the programming education they require~\cite{koutropoulos2012emotive}, they provide significantly less control over the student and their progress~\cite{kinash2013moocing}, which means that the student may pass the course without learning everything they need to know. 
Specifically, it is possible to submit solutions to programming tasks without maintaining the desired level of code quality, because, on the one hand, it is not feasible to manually check this many submissions~\cite{carter2003shall}, and on the other, professional automated code quality tools are not adapted to the education process~\cite{keuning2019teachers}.

Much research has been conducted that studies the aspect of code quality in code written by students in MOOCs or regular classes~\cite{keuning2017code, de2018understanding, edwards2017investigating, albluwi2020using, aivaloglou2016kids, techapalokul2017understanding, bai2019amelioration}.
The majority of the existing research focuses on Java~\cite{keuning2017code, de2018understanding, edwards2017investigating, albluwi2020using} and Scratch~\cite{aivaloglou2016kids, techapalokul2017understanding}, since public large-scale datasets already exist for these languages~\cite{brown2014blackbox, scratchOpenDataset}.
Despite the growing popularity of Python in education~\cite{lo2015programming}, Python code quality studies are usually limited to submissions from just one or several semesters of a university course~\cite{liu2019static, albluwi2020using}.
Moreover, researchers usually only consider code fragments of tasks for beginners~\cite{keuning2017code, aivaloglou2016kids, techapalokul2017understanding, albluwi2020using}, whereas it would also be valuable to look at more complex code snippets.
Finally, existing works do not take into account external factors of code quality issues, \eg the issues in the task itself, which can affect the results.

To bridge the existing gaps in research, in this work, we conduct a large-scale analysis of Java and Python student submissions from a popular MOOC platform --- JetBrains Academy~\cite{jetbrainsAcademy}.
The platform's team provided us with a large dataset of \javasubmissions Java submissions and \pythonsubmissions Python submissions for more than 700 different tasks.
This includes only the correct submissions, after passing all tests, allowing us to study code quality issues in them.
We applied a tool called Hyperstyle~\cite{birillo2021hyperstyle} to discover code quality issues in the submissions. 
Hyperstyle is embedded into JetBrains Academy, meaning that students see its feedback after a successful task submission and can fix detected code quality issues in their next attempts to increase their code quality grade.
This allowed us to use the historical data of past submissions to analyze the way students resolve existing issues.

The analysis of the most popular code quality issues showed that among both languages, not only minor issues like unused imports are popular, but also issues that can lead to bugs in the future, \textit{e.g.}, the incorrect usage of a switch statement in Java or the shadowing of built-in function names in Python.
This indicates the necessity to teach students about them early on.
Moreover, we found that frequent code quality issues are not always the fault of students, and can be caused by tasks on the platform itself: for example, incomplete or incorrect theory part, a weak test base, mistakes in task descriptions or in pre-written code templates, an incorrect order of topics that leads to the misunderstanding of the learned concepts, etc.

Next, we looked at the way students fix their issues.
We found out that the majority of students are good at fixing simple issues that do not require big changes in the code.
Moreover, some students even tried to correct code quality issues that were not their fault, \textit{e.g.}, fix issues in a pre-written template or in a code sample taken from an incorrect theory part.
This indicates a successful synthesis of the code quality analyzer and the platform, since students try to follow the advice of the code quality analyzer to fix issues even if they did not introduce them.
During the detailed manual analysis of the submissions, we discovered several interesting cases, \eg many students try to shorten their code and only worsen its quality due to the lack of experience in refactoring.

Finally, we analyzed abnormally long sequences, where after successfully passing the task, students submit five or more attempts, changing or fixing the code.
It turned out that students can spend dozens of attempts improving their code and get the highest code quality score. 
We noticed that the main reason for many attempts is the student's lack of understanding of what exactly needs to be corrected. 
For example, in the code editor on the platform, only the line with the issue is highlighted, but not the exact position, which confuses students, so they try different fixes until they reach the correct code.
Another interesting observation is that students correct issues by groups, \textit{i.e.}, if students see a familiar issue and know how to correct it, they try to correct it in all the places where it appears at once.

The results of our study and the provided insights can be of use to various groups of practitioners. 
Teachers who are developing or maintaining programming courses can use our results to modify their content to focus on the most prevalent issues. 
Software developers who are creating new code quality tools can consider our insights to take care of the most prominent mistakes, as well as to facilitate their better fixing among students.
Finally, content managers and creators of MOOCs and their tasks can ensure that the content itself does not facilitate the issues, be that in the theory or templates.
 
The tool that we developed for gathering data from JetBrains Academy is available online, among with the detailed statistics about the code quality issues within our dataset~\cite{artifacts}, so others can extend this paper.

Overall, the main contributions of our paper are:

\begin{enumerate}
    \item \textbf{Analysis} of more than a million successful submissions in Java and more than 1.3 million successful submissions in Python from the standpoint of their quality. We describe the most popular types of code quality issues and analyze the dynamics of students fixing them. 
    \item \textbf{Case studies} of the most prominent issues, where we provide several examples that demonstrate how students struggle with code quality problems and how a MOOC's environment can help or interfere with the studying process.
    \item \textbf{Insights} and practical implications that can be useful to teachers who are in the process of developing courses, as well as developers working on MOOC platforms and their quality control tools.
\end{enumerate}

\section{Background}\label{sec:background}

\subsection{Code Quality Analysis}
Many works study the problems of students' code quality, analyzing large-scale datasets of student submissions to find certain common behaviors and patterns~\cite{edwards2017investigating, albluwi2020using, aivaloglou2016kids, techapalokul2017understanding, keuning2017code, effenberger2022code} and testing code quality tools on small groups of students~\cite{molnar2020using, liu2019static, keuning2020student}.
The first group mostly focuses on Scratch and Java languages, and uses open datasets for their research~\cite{brown2014blackbox, scratchOpenDataset}.

Keuning et al.~\cite{keuning2017code} studied Java submissions from the Blackbox database~\cite{brown2014blackbox}. 
The authors used PMD~\cite{pmd}, a pre-configured industrial code analyzer, to find common code quality issues. In addition, the authors studied which issues students correct over time, and compared code quality of students who use code analyzers with those who do not.
The main weak point of this work is the lack of studying issues of more experienced students~\cite{edwards2017investigating}.

Edwards et al.~\cite{edwards2017investigating} broadened the research towards more difficult tasks. They studied over 500,000 Java submissions from students with different programming experience, and categorized the obtained code quality issues, especially of formatting issues.
The students used the Web-CAT tool~\cite{edwards2008web} to submit their solutions and immediately receive feedback from professional analyzers.
In addition to researching general issues, the authors explored the issues with more granular metrics, for example, which issues take more time to fix.
The authors found that students correct issues poorly since professional linters are not adapted to the educational process, \ie provide feedback that is difficult for students to understand~\cite{edwards2017investigating, keuning2019teachers}.

Albluwi et al.~\cite{albluwi2020using} studied the distribution of code quality issues from professional linters. 
The authors searched for a correlation between the issues and various metrics, such as student experience and progress in the course, course difficulty, etc., but the size of the dataset they used was rather small. 
The data includes submissions of 968 students from three semesters of an introductory programming course 
that solve nine programming tasks, which might be too small to generalize the results.
The above-mentioned problem of having unadapted feedback is also present.

The next group of works focused on Python. 
The \textsc{PyTA} tool was developed by Liu and Petersen~\cite{liu2019static} to provide custom checks for common code quality issues with adapted feedback. 
This tool is a wrapper for the popular Pylint~\cite{pylint} analyzer.
The authors compared the issues before and after embedding the tool into a computer science course, and found that students who use the tool make fewer mistakes and spend less time fixing them.
The data consists of 114,865 submissions from 40 coding exercises, however, there is no detailed analysis of the code quality issues in these solutions, and the work is aimed only at the beginner students.

Molnar et al.~\cite{molnar2020using} analyzed Python code in submissions of various difficulty using Pylint~\cite{pylint} and employed their own tool for visualizing the issues.
The authors used a special score computed by Pylint to assess the quality of the code.
The work shows how the code quality score changes depending on the complexity of the tasks, as well as depending on the progress of the given student.
However, the dataset only contains 642 submissions, and, similarly to previous works, an unadapted code quality tool was used for the assessment.

Effenberger and Pelánek~\cite{effenberger2022code} recently analyzed code quality issues in 114,000 Python submissions for 161 tasks, taken from 11,000 students over 2.5 years.
The authors used the data for tasks that can be solved with 2--20 lines of code with a median time for their solving from 1 to 20 minutes.
The authors analyzed not only the issues that exist in correct solutions, but also how the provided feedback affects their correction. However, the main limitation of this work is the simplicity of the tasks.

Overall, it can be seen that the topic of the quality of student code is important and rather well-researched, however, existing studies share a number of common drawbacks. 
The open-access datasets (\textit{e.g.}, Blackbox~\cite{brown2014blackbox}) are large, but they do not contain detailed information about the student's full context, like the particular task the student is working on~\cite{keuning2017code}.
Other datasets are significantly smaller---from several hundred~\cite{molnar2020using} to a hundred thousand submissions~\cite{liu2019static}---and only contain tasks for novices.
Finally, researchers use tools, the output of which is not adapted to the educational process, and students may find it difficult to interpret~\cite{keuning2019teachers}.

In our work, we aim to overcome these limitations by conducting a study using a large dataset of solutions from a popular education platform JetBrains Academy~\cite{jetbrainsAcademy} that employs a code quality tool called Hyperstyle~\cite{birillo2021hyperstyle} within it.

\subsection{JetBrains Academy and Hyperstyle}\label{sec:background_platforms}

\begin{figure*}[t]
    \centering
    \includegraphics[width=0.8\linewidth]{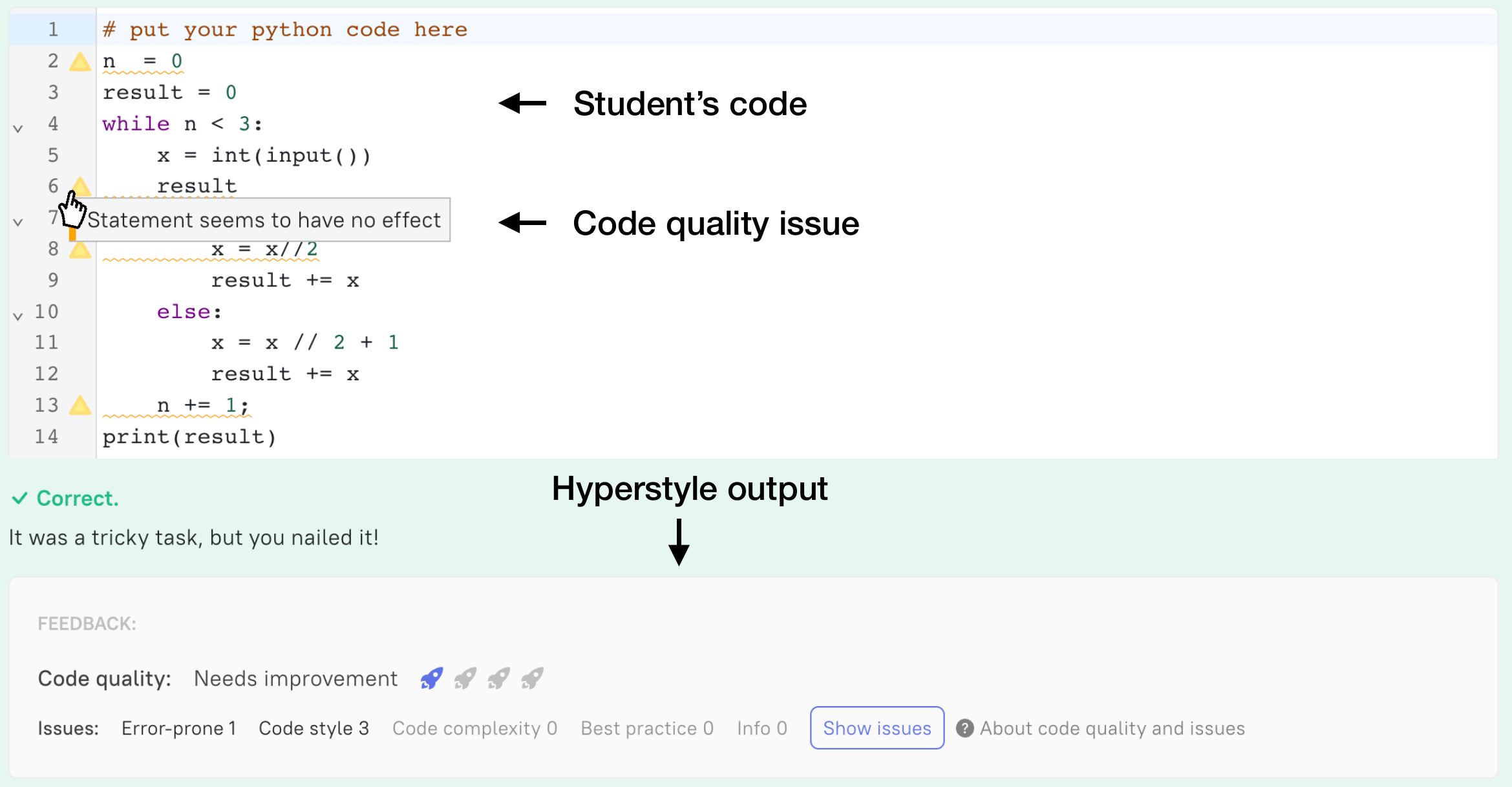}
    \caption{The UI of a task on JetBrains Academy, together with a Hyperstyle issue tooltip in the editor and the tool's output.}
    \label{fig:hyperstyle}
    \vspace{-0.1cm}
\end{figure*}

JetBrains Academy~\cite{jetbrainsAcademy} is a project-based education platform developed by JetBrains~\cite{jetbrains}.
The learning on the platform is organized as completing a chain of small tasks that are all part of a single complex one, thus, after solving them, the student has a finished project.
Currently, the platform supports studying Java, Python, Kotlin, JavaScript, and Go. The platform can be used by students, but is also available for everybody. Some courses are free, some require a paid subscription, both Java and Python courses that are the focus of this study are paid.

JetBrains Academy is built around a large knowledge graph, reflecting the set of topics necessary for studying a programming language.
The topics can be both theoretical (\eg asymptotic complexity) and practical (\eg arrays) with some topics being common between languages (\eg algorithms or data structures).
To learn more complex concepts, it is necessary to move in deeper along the graph.
Each task in a studied project is connected to one topic in the graph, aligning the development of the project with the learning of concepts.
This allows defining the \textit{complexity} of the task, \textit{i.e.}, the depth in the knowledge graph of the topic that corresponds to this task.
There are three levels of complexity depending on the depth: \textit{shallow} ($\le 2$), \textit{deep} ($\ge 5$), otherwise \textit{moderate}.

Submissions to programming tasks are validated in two stages. 
Firstly, the correctness of the solution is checked using tests, similar to most MOOCs. 
If all the tests pass, the second stage is initiated where code quality is checked using the Hyperstyle~\cite{birillo2021hyperstyle} tool.
Hyperstyle uses a subset of code quality checks of professional analyzers (like PMD~\cite{pmd} for Java or Pylint~\cite{pylint} for Python) 
that are relevant to the educational process. Also, it changes the messages for the majority of the analyzers' issues to be more understandable for students.
As a result, students receive a grade based on the overall quality of their code.
Different code quality issues affect the final grade differently, some issues are only informative (\textit{minor} issues) and do not lower the grade at all.

The detected issues are also highlighted in the JetBrains Academy platform's user interface, so that the students can see and fix them in subsequent attempts and improve their grades. The UI of JetBrains Academy, together with the Hyperstyle issue and output, are presented in Figure~\ref{fig:hyperstyle}. It is also important to mention that students who have successfully passed the task can see other public submissions of fellow students with highlighted code quality issues. 

All of this makes JetBrains Academy a suitable platform to study code quality issues. It already comes with the embedded code quality tool, which allows us not only to analyze the issues themselves, but also check the history of submissions from students to track how they fixed these issues.
\section{Dataset}\label{sec:dataset}

For our analysis, the JetBrains Academy team provided the data about students' successful submissions on the platform over one year (from September 1st, 2020 to September 1st, 2021) that are written in Java or Python and passed all the correctness tests for the corresponding task.
Every submission contains the source code of the solution, various metadata like timestamp, task ID, user ID, etc., as well as platform-specific features 
such as the task's complexity. 
All personal information about users was completely removed by the platform's team. 
The obtained data was preprocessed and then filtered.

\textbf{Preprocessing.}
All successful solutions submitted by each user to a particular task consecutively were grouped and sorted by timestamp, we will refer to such a collection as a \textit{submission series}. 
If two consecutive submissions in one series were identical, the latter was filtered out.
Next, for each submission, we ran the Hyperstyle tool and collected its output to see what code quality issues are present in it. We use the output of the Hyperstyle tool, since this tool is used on the platform to show code quality issues to the students.

\textbf{Filtering.}
Firstly, we omitted issues found in \textit{templates} --- pre-written parts of the solution from the creators that students can edit and supplement with their own code.
Issues in the template were introduced by the task creators, not students, so we did not take them into account in our analysis.
To filter them, we applied an algorithm that is used on the platform itself to find code quality issues in the templates and fix them.
The algorithm analyzes students' solutions, looks for issues that are most frequently not corrected, matches positions with the template, and finally marks them as template issues.

In order to select more representative and interesting data for the general analysis, we applied a number of filtrations. 
Firstly, for the main analysis we kept only the submission series that contain five successful submissions or fewer. 
This allowed us to remove suspiciously long series but filter out only 5\% of the data, since the vast majority of users finish the task in under five attempts.
Secondly, we skipped all formatting issues like incorrect indentation or missing whitespaces to focus on more prominent and serious issues.
Although formatting issues are important and appear in student's solution more often, they are already well-studied~\cite{edwards2017investigating, albluwi2020using, liu2019static} and the process of correcting them is rather straightforward.
The characteristics of the final dataset are presented in \Cref{tab:dataset}.

\begin{table}[t]
\centering
\caption{Characteristics of the collected dataset \\ after preprocessing and filtering.}
\begin{tabular}{lrr}
\toprule
                            & \textbf{Java}   & \textbf{Python}   \\ \midrule
\textbf{Tasks}              & \javasteps       & \pythonsteps       \\
\textbf{Students}           & \javastudents    & \pythonstudents    \\\midrule
\textbf{Solutions to shallow tasks}  & \javashallow     & \pythonshallow     \\
\textbf{Solutions to moderate tasks} & \javamiddle      & \pythonmiddle      \\
\textbf{Solutions to deep tasks}     & \javadeep        & \pythondeep        \\ \midrule
\textbf{Total solutions}    & \javasubmissions & \pythonsubmissions \\\bottomrule
\end{tabular}
\label{tab:dataset}
\vspace{-0.4cm}
\end{table}

\section{Analysis}\label{sec:analysis}

In our analysis, we studied the following two research questions about the code quality of student submissions.

\begin{enumerate}[leftmargin=1.3cm,start=1,label={\bfseries RQ\arabic*:}]
    \item Which code quality issues are the most prevalent in Java and in Python submissions?
    \item How do students fix various types of issues as they update their submissions with further attempts?
\end{enumerate}

\subsection{\textbf{RQ1: Most Prevalent Issues}}
In this section, we study the most frequent issues and the reasons for their popularity.

\textbf{Methodology.}
In each submission series, we selected only the first attempt, since we wanted to focus on the students' initial issues, before they had seen any feedback from the linters.
These attempts represented \javafirstattemptper\% of all submissions for Java and \pythonfirstattemptper\% for Python, respectively.
For each issue type, we calculated the percentage of submissions in which it was detected at least once, separately for the tasks with different complexity levels.
Finally, we compared the most prevalent issues and the differences between complexities.

\begin{figure}[t]
    \centering
    \includegraphics[width=\columnwidth]{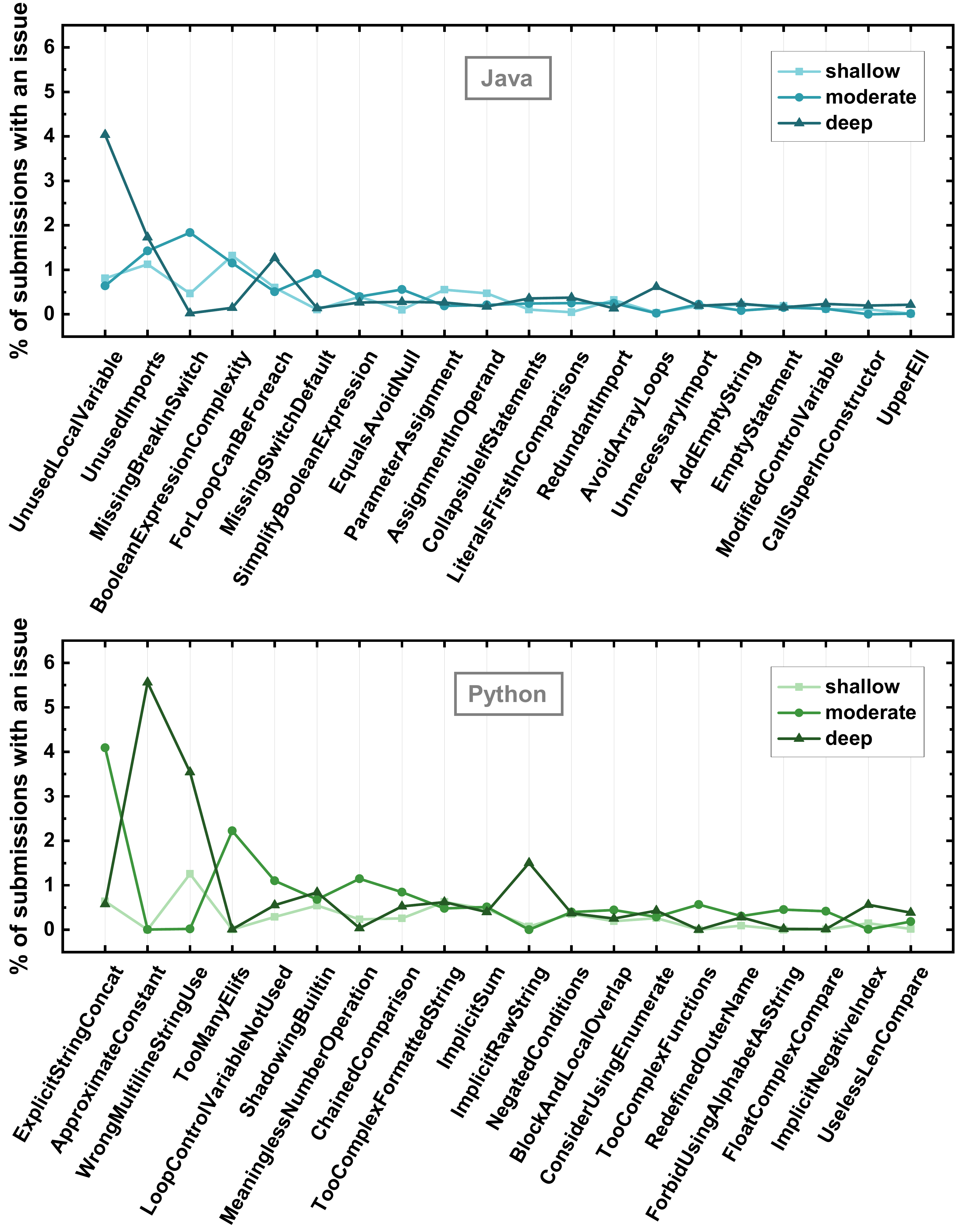}
    \vspace{-0.5cm}
    \caption{Distribution of the most prevalent issues among Java submissions (top) and Python submissions (bottom). The issues are sorted by their total prevalence in all submissions.}
    \label{fig:issues}
    \vspace{-0.5cm}
\end{figure}

\textbf{Results.} Firstly, it is important to note that only 8\% of Java and 11\% of Python first attempt submissions have at least one code quality issue not connected to formatting.
The percentages for each issue type are presented in \Cref{fig:issues}, the top part shows the most popular issues in Java, while the bottom part shows the information for Python. Both parts are sorted by the total percentage among all submissions.

The Top-5 Java issues include: \textbf{(1)} \textit{Unused local variable}, which finds a locally declared variable that is not used; \textbf{(2)} \textit{Unused imports}, which is similar in regards to imports; \textbf{(3)} \textit{Missing break in switch}, which points to a possibly skipped \texttt{break} operator in a casing construction; \textbf{(4)} \textit{Boolean expression complexity}, which warns against conditions that are too complex, \textit{e.g.}, \texttt{if (a > -15 \&\& a <= 12 || a > 14 \&\& a < 17 || a >= 19) \{...\}}; and \textbf{(5)} \textit{For loop can be foreach}, which advises to rewrite the \texttt{for} loop into the \texttt{foreach} to avoid explicit indexing.

The Top-5 Python issues include: \textbf{(1)} \textit{Explicit string concatenation}, which forbids using string concatenation; \textbf{(2)} \textit{Approximate constant}, which finds the constants that the students defined and that can be replaced by existing constants from the \texttt{math} library; \textbf{(3)} \textit{Wrong multiline string use}, which informs about the incorrect usage of multiline strings as function arguments;  \textbf{(4)} \textit{Too many elifs}, which informs about using \texttt{elif} statements that are too complex;  and \textbf{(5)} \textit{Loop control variable not used}, which finds unused variables in loops that should be replaced with the underscore symbol.
 
While there exists a difference between different task complexities, we cannot definitely conclude that some issues are more popular in moderate tasks than in deep ones. Usually, the high frequency of the issue can be explained by the task itself. For example, \textit{Missing break in switch} is more popular in moderate tasks, because students learn this construct in a task with such complexity.
The same findings were highlighted by Effenberger and Pelánek~\cite{effenberger2022code}. 
They found that some issues are highly localized in several tasks and thus influence the general distribution.
In Section~\ref{sec:case:studies}, we touch upon some of them in more detail and give concrete code examples to illustrate the most peculiar situations.

\subsection{\textbf{RQ2: Dynamics of the Solution Quality}}
In this section, we study how students fix quality issues in their submission series.
While the Hyperstyle paper~\cite{birillo2021hyperstyle} has studied the influence of the tool on the correction of code quality issues, a detailed analysis of which issues get corrected was not provided.

\textbf{Methodology.} 
Firstly, we filter out all submissions with only a single attempt, which is 81\% of all submission series for Java and 87\% for Python. 
Of these, less than 10\% have at least one code quality issue, however, the students chose not to fix them.
Next, for each submission series that we kept, we calculate the percentage of issues left in the last submission in the series relative to the first attempt.
For example, if there were 1,000 initial submissions with the given issue, and 200 in the last attempts, we say that the issue remained in 20\% of cases.
Comparing the first and last attempts allows us to focus on the high-level, overall nature of fixing issues.

\begin{figure}[t]
    \centering
    \includegraphics[width=\columnwidth]{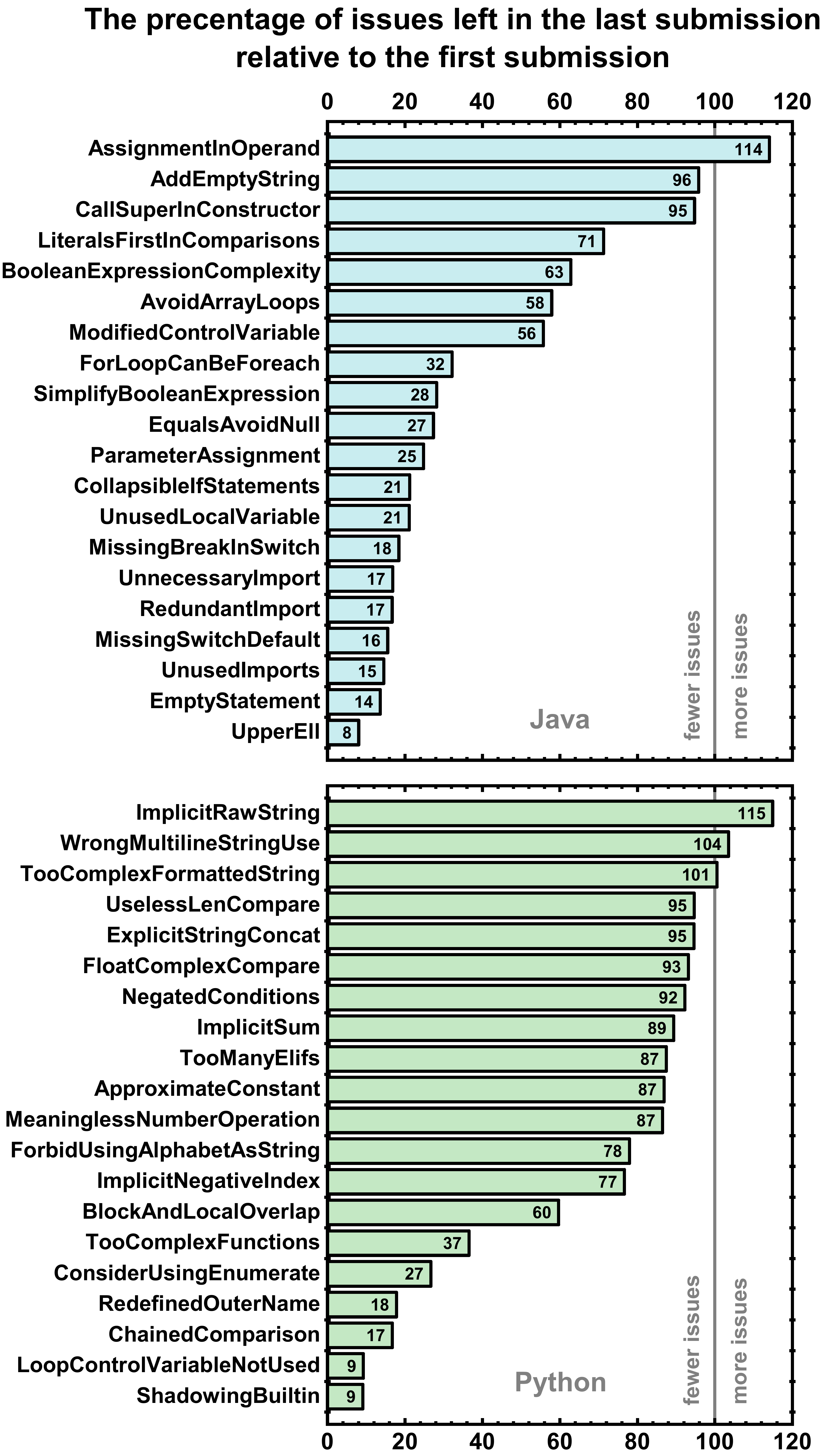}
    \vspace{-0.5cm}
    \caption{The percentage of submission series that have each issue left in the last attempt, for Java (top) and Python (bottom). Lower values indicate better fixing, higher values indicate worse fixing.}
    \label{fig:dynamic}
    \vspace{-0.5cm}
\end{figure}

\textbf{Results.}
The dynamics of fixing various issues in submission series is shown in \Cref{fig:dynamic}, the top chart represents Java submissions, while the bottom one represents Python submissions. 
In the chart, we consider the most prevalent issues that were identified in RQ1.

For Java, many issues are successfully corrected by students and remain in the final submission in less than 15\% of cases: \textbf{(1)} \textit{Upper ell} forbids the student to use a lower-case \texttt{l} in numbers of the \texttt{long} type, since the lower-case \texttt{l} looks a lot like \texttt{1}; \textbf{(2)} \textit{Empty statement} informs the student about standalone ``;'' semicolon with no statement; and \textbf{(3)} \textit{Unused imports}.
These issues are easy to recognize and understand.
Moreover, to correct such issues, students need to add or delete a single line or keyword, which explains the fact that students successfully fix them.
However, several issues are not corrected in more than 90\% of cases: \textbf{(1)} \textit{Call super in constructor} reminds of the need to call the constructor of the class of the child; \textbf{(2)} \textit{Add empty string} indicates a useless concatenation with an empty string.
To make matters worse, there is a 14\% increase in submissions with the \textit{Assignment in operand} issue, which prohibits an assignment within the conditional expressions of \texttt{while} or \texttt{if} statements. 
We examine the reasons for this anomaly in detail in \Cref{sec:rq2:case:study}. Keuning et al.~\cite{keuning2017code} also showed that some mistakes are not corrected by students, while some are corrected by the majority.

For Python, there are only two issues which remain in less than 15\% of students' submissions: \textbf{(1)} \textit{Shadowing Python built-in}, which appears if students override built-in functions like \texttt{max} or \texttt{input}; and \textbf{(2)} \textit{Loop control variable not used} when loop control variable is not in use and can be replaced with an underscore.
Moreover, unlike Java, most of the issues in the studied list are not corrected by students even in the final attempt in more than 50\% of cases.

Finally, as many as three of them appear more often in the last attempt than in the first: \textbf{(1)} \textit{Too complex formatted string}, which is connected to using  \texttt{f-strings} incorrectly (strings like \texttt{f"My name is: \{name\}"}); \textbf{(2)} \textit{Wrong multiline string use}, which forbids using multiline strings directly; \textbf{(3)} \textit{Implicit raw string} which forbids escape sequences inside regular strings and forces to use Python raw strings instead.
Similar to Java, we analyzed these anomalies separately in the case studies in \Cref{sec:rq2:case:study}, and came to the conclusion that in all cases it is not the fault of the students. 
The reason lies in the platform itself, namely, in the incomplete or erroneous theoretical part or the task idea and formulation in general, which lead to the inability to implement the task correctly.

\section{Case studies}\label{sec:case:studies}

Generally, it is difficult to find the true root cause of the popularity of an issue without manually studying specific cases.
In this section, we describe several cases that provide explanations for some major peaks in our data.

\subsection{\textbf{RQ1: Most Prevalent Issues}}

\textbf{Method.}
For each code quality issue with high frequency (from \Cref{fig:issues}), we analyzed its distribution among different tasks.
We assume that if the distribution is biased towards a specific task, there should be some problem with the task itself.
By examining the reason for this accumulation of issues in a particular task with concrete examples from the dataset, we can identify the true reason why students make this issue so often and whether it is their own fault or not.
As examples, we selected and analyzed around 20 submissions from a concrete task that contains such an issue and the same amount where students avoid making it.

\begin{figure}[t]
    \centering
    \includegraphics[width=\columnwidth]{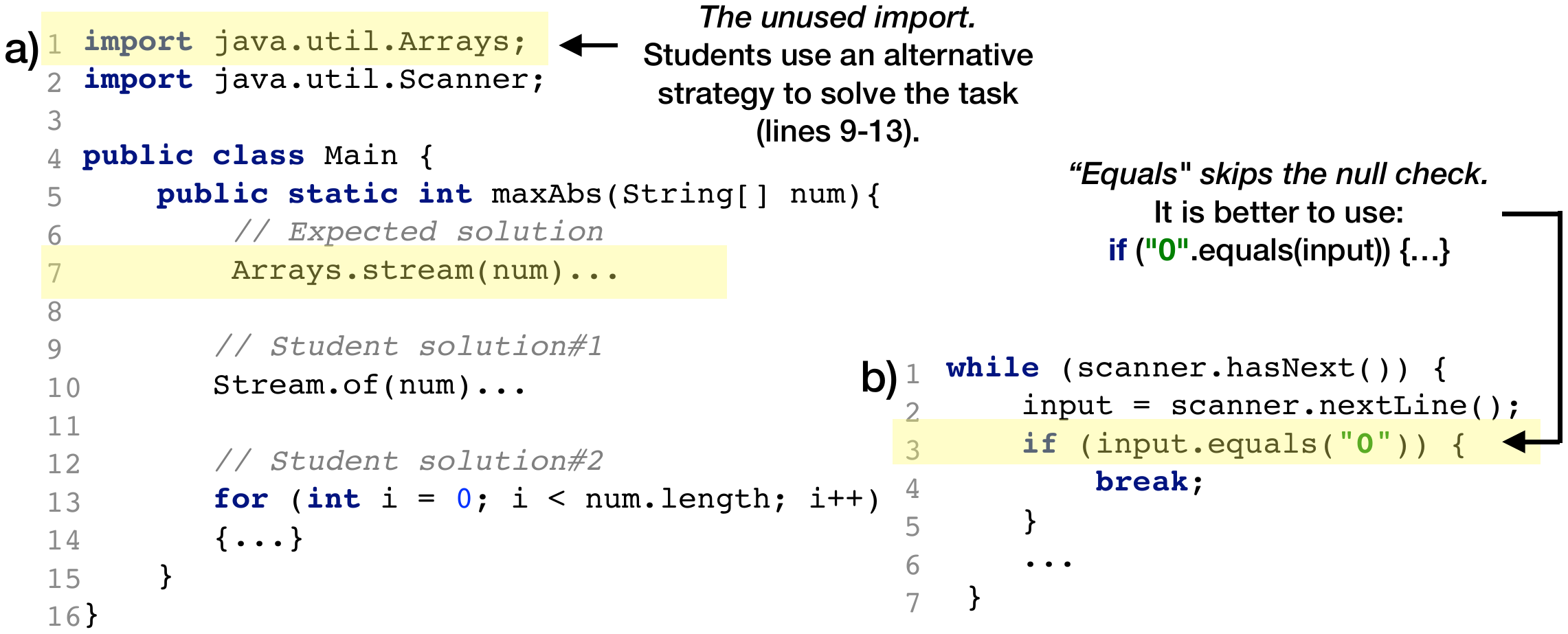}
    \vspace{-0.5cm}
    \caption{Examples of RQ1 Java cases: a) Unused imports; b) Equals avoid null. The code formatting is unchanged.}
    \label{fig:rq1:cases:java}
    \vspace{-0.5cm}
\end{figure}

\textbf{Case studies.}
In total, we highlight six cases: three for Java (\Cref{fig:rq1:cases:java}) and three for Python (\Cref{fig:rq1:cases:python}).
Each case demonstrates how different factors, \textit{e.g.}, incorrect theory part, irrelevant tests, or complexity can influence code quality.

\subsubsection{\textbf{Unused local variable} (Top-1 Java issue)}

This issue appears in 80\% of the student submissions for Task\#9057~\cite{task9057}, which asks to read several values, put them into placeholders in the text, and print the result.
However, not all input values are required to complete the text, so if the student uses a separate variable for each of them, some remain unused.
Some students avoid this issue by reading the value without saving it into a variable, but apparently, most of the students do not come to this idea because of the lack of experience in programming.
We reported this observation to the creators of the task, now all the variables are used, and students no longer have such an issue in their submissions.

\textit{Summary.}
This example shows that the massive prevalence of this code quality issue was caused by poorly designed and tested tasks combined with the lack of knowledge by the students.
Such kind of ``third-party'' mistakes, not related to the main idea of the task, can discourage the student and interfere with their learning of new complex material.
Therefore, one possible solution to avoid this problem in this particular task would be to display this inspection without lowering the score, as optional or additional information, or even disable it.

\subsubsection{\textbf{Unused imports} (Top-2 Java issue)}
This issue is common for all tasks, and most of it happens by accident because of the students' carelessness.
However, one example we would like to highlight is Task\#3828~\cite{task3828}, where students need to find the maximum value in an array of numbers (\Cref{fig:rq1:cases:java}a).
The task assignment explicitly says: \textit{``Try not to use the loop, but use Stream API''}.
Moreover, the template for the task contains import of \texttt{java.util.Arrays} to work with the Stream API.
However, some students ignore this and write the code using a \texttt{for} loop, probably because they are more familiar with this construct. 
Conversely, more advanced students use \texttt{java.util.stream.Stream} library as an alternative for one mentioned above.

\textit{Summary.} This is an example of how a code quality issue can indicate different intentions of teachers and students.
The authors of the tasks expected only one particular solution from the students and configured the template only for that solution.
However, they did not consider that some tasks can have alternative solutions, in which some parts of the template may not be needed, and therefore, cause an issue.
In order to better control the submissions, the test base for these tasks should be strengthened or even extended with semantic checks.

\subsubsection{\textbf{Equals avoid null}} This issue is interesting because it dominates in tasks with \textit{moderate} complexity, although the peaks should logically occur in \textit{shallow} tasks due to the inexperience of students.
The cause of this peak is a specific \textit{moderate} task, where 40\% of all submissions with such an issue are concentrated.
It requires the student to use the \texttt{equals} method to compare the input string with the constant string ``0''.
However, the input value in Java can be \texttt{null}, so to avoid a \texttt{NullPointerException}, the best practice is to invoke \texttt{equals} on the constant, passing the input as an argument, rather than vice versa (see \Cref{fig:rq1:cases:java}b). 

\textit{Summary.} This task contains a theoretical section, the theory only provides examples of comparing two variables, so the student can find out about the practice of variable-to-constant comparison only from the Hyperstyle's feedback.

\begin{figure}[t]
    \centering
    \includegraphics[width=0.9\columnwidth]{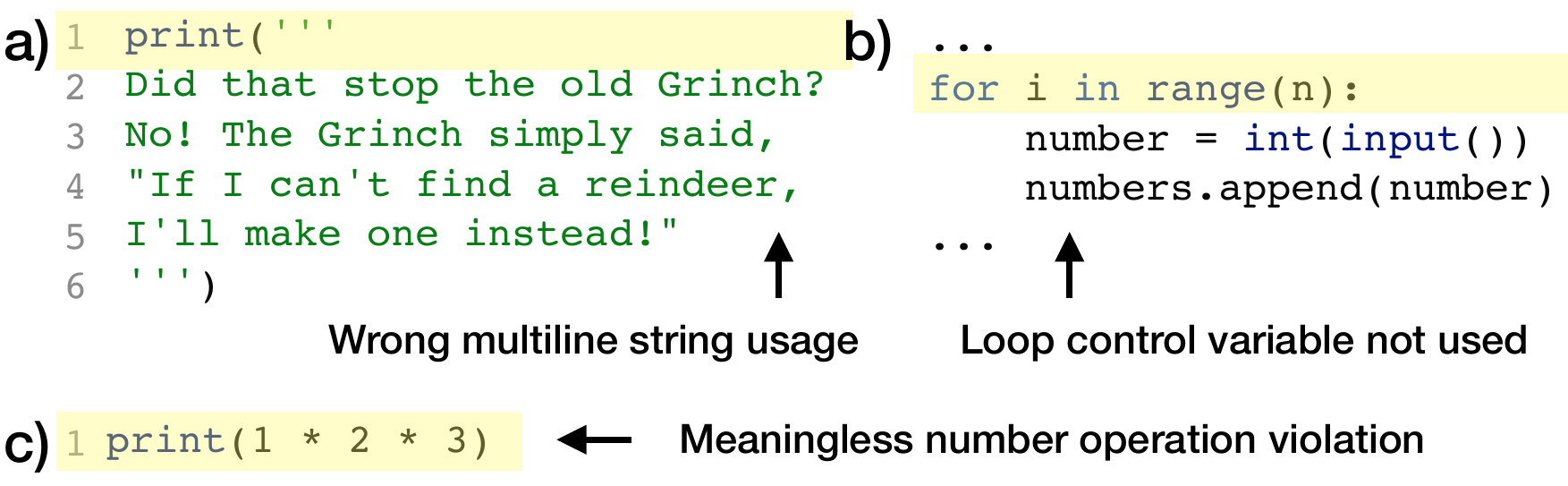}
    \vspace{-0.1cm}
    \caption{Examples of RQ1 Python cases: a) Wrong multiline string use; b) Loop control variable not used; c) Meaningless number operation violation. The code formatting is unchanged.}
    \label{fig:rq1:cases:python}
    \vspace{-0.5cm}
\end{figure}

\subsubsection{\textbf{Wrong multi-line string use} (Top-3 Python issue)}
The prevalence of this issue is mostly caused by
Task\#6881~\cite{task6881}, which requires students to print a multi-line string. 
In the first attempt, 88\% of the students sent the straightforward submission with the multi-line string directly inside the \texttt{print} function (\Cref{fig:rq1:cases:python}a), which is flagged by Hyperstyle.
Only a few students managed to avoid this issue by submitting the following solutions: \textbf{(1)} divide the string into several \texttt{print} calls; \textbf{(2)} use a single-line string with \texttt{\textbackslash n} newline separators and invoke \texttt{print} only once; \textbf{(3)} put the text into a variable. 

\textit{Summary.} This task also contains a theoretical section. 
The theory part has a multi-line string usage example that contradicts the official Python style guide.
When students try to use the incorrect code from the example, the issue is flagged.

\subsubsection{\textbf{Loop control variable not used} (Top-5 Python issue)}
Task\#6818~\cite{task6818} requires students to write a program that calculates the arithmetic mean. 
To solve the task, students add a loop without using its counter. In these cases, the variable should be replaced by an \texttt{\_} symbol (see \Cref{fig:rq1:cases:python}b).

\textit{Summary.} The theoretical section of this task does not have an example of this language feature.
As a result, 48\% of student submissions contain the issue in the first attempt.
However, thanks to Hyperstyle, the students learn about it and in more than 90\% of the cases get rid of it during their following attempts.

\subsubsection{\textbf{Meaningless number operation}}
This is another example of an issue where the peak in \Cref{fig:issues} can be seen for the \textit{moderate} tasks, which can be once again explained by specific tasks with moderate complexity causing it.
Task\#6558~\cite{task6558} requires students to write a program that prints the product of these three numbers: \texttt{1 * 2 * 3} (see \Cref{fig:rq1:cases:python}c). However, according to the linter's rules, it is meaningless and could be simplified into \texttt{print(6)}.
As a result, about 85\% of these issues related to this particular task. 

\textit{Summary.} Obviously, the authors tried to artificially simplify the task for beginner students, missing the possible code quality issues.
For this task, it would be better to use values from the input instead of the predefined constants, or put the code for obtaining input inside the pre-written template.
This could probably keep the student’s task simple but eliminate code quality issues.
An alternative solution here could be to hide this issue from the student by customizing Hyperstyle for this particular task.

 \textbf{Discovered causes of code quality issues}. \textbf{(1)} The task assignment is poorly developed and tested;
 \textbf{(2)} Students do not use constructs that are proposed by the creators;
 \textbf{(3)} The theoretical part of the task is incomplete or even incorrect;
 \textbf{(4)} Students have a lack of knowledge to avoid the issue.

\vspace{0.2cm}
\subsection{\textbf{RQ2: Dynamics of the Solution Quality}}~\label{sec:rq2:case:study}

\vspace{-0.2cm}

\textbf{Method.}
To explain the differences in the dynamics of fixing issues, we also conducted a manual review of submissions, and collected some illustrative examples.
For each issue, we were interested in the submission series of three different configurations: \textbf{(1)} only the first attempt contains an issue (the issue was fixed), \textbf{(2)} both the first and the last attempts contain an issue (the issue was not corrected), and \textbf{(3)} only the last attempt contains an issue (the issue appeared in the subsequent solutions).
For this analysis, we selected tasks with the highest frequency of issues and analyzed their submission series.

\begin{figure}[t]
    \centering
    \includegraphics[width=\columnwidth]{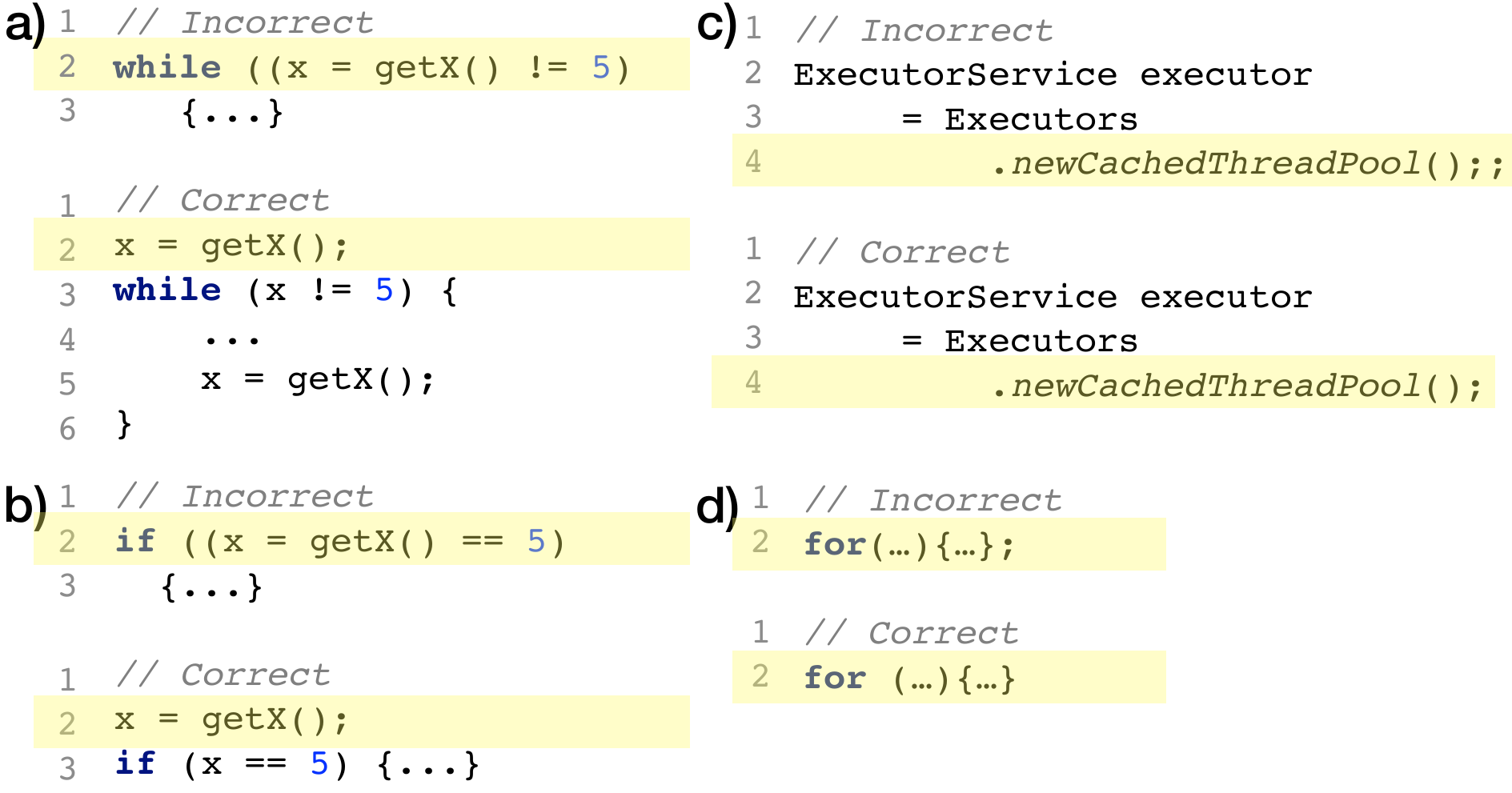}
    \caption{Examples of RQ2 Java cases: a, b) Assignment in operand; c, d) Empty statement check. }
    \label{fig:rq2:cases:java}
    \vspace{-0.2cm}
\end{figure}

\textbf{Case studies.}
We considered four different cases with examples and explanations for several interesting trends, which we observed in \Cref{sec:analysis}.
\Cref{fig:rq2:cases:java} shows examples for the Java cases and \Cref{fig:rq2:cases:python} --- for the Python cases.

\subsubsection{\textbf{Assignment in operand} (Top-1 Java unfixed issue)}

This issue often occurs in tasks where students are required to read numbers until a specific number is seen, \eg Task\#2153~\cite{task2153}.
There are several ways to implement this, but the most popular among students is to read the first number, assign it to the variable, and then read the remaining numbers into the same variable in the loop, until it equals the necessary value (see \Cref{fig:rq2:cases:java}a).
To get rid of code duplication, experienced students start to save the numbers into a variable directly inside the \texttt{while} condition (\Cref{fig:rq2:cases:java}a and \Cref{fig:rq2:cases:java}b), which leads to an \textit{Assignment in operand} issue.
It is worth noting that this code is also correct and is even sometimes used for size optimization of the bytecode. 
However, in industrial programming, this code construction is considered to be error-prone, as it is difficult to read and understand.

\textit{Summary.}
The theory on the platform does not provide such a complex example of the \texttt{while} usage.
However, we found many students that have only two submissions with a small difference in time (within 5 minutes) and the second one has this complicated case with the \texttt{while} usage.
Taking all facts into account, we can assume that these students, after successfully submitting a solution, used the opportunity to see other students' successful public solutions on the JetBrains Academy platform, saw this interesting new language construct with the \texttt{while} usage, copied it into their code and submitted it.
This example shows how a seemingly useful platform feature can have an unexpected effect in the form of spreading a mistake.

\subsubsection{\textbf{Empty statement check} (Top-2 Java fixed issue)}
We found three typical instances of this issue. 
Firstly, there are two semicolons one after another next to a variable initialization. The task has a pre-written template with a variable declaration with a semicolon at the end. 
When students initialize the variable with a value, they might also add their own semicolon automatically, forgetting about the one already present in the template (see \Cref{fig:rq2:cases:java}c).

Secondly, there can be a redundant semicolon after a control flow statement. Students are taught that in Java it is necessary to put a semicolon after each statement, so diligent students also do this after the body of \texttt{if}, \texttt{for}, and \texttt{while} blocks (\Cref{fig:rq2:cases:java}d), although it is not required.

Finally, there can be a control flow statement without a body like \texttt{for (i = 2; i <= n; i *= 2);}.
Oftentimes, students try to shorten their code and calculate some value directly using a \texttt{for} loop variable, leaving the body of the \texttt{for} statement empty. For example, this can happen to calculate the power of some value.
However, this is an anti-pattern, since an empty \texttt{for} body often introduces bugs that are hard to spot later on.

\textit{Summary.}
Most students (over 85\%) fix this issue because it seems pretty easy to fix --- simply remove the extra semicolon. 
However, about 15\% of users do not to fix this issue. 
A possible reason for this may be that the issue is too simple and students prefer to fix more interesting ones.

\begin{figure}[t]
    \centering
    \includegraphics[width=\columnwidth]{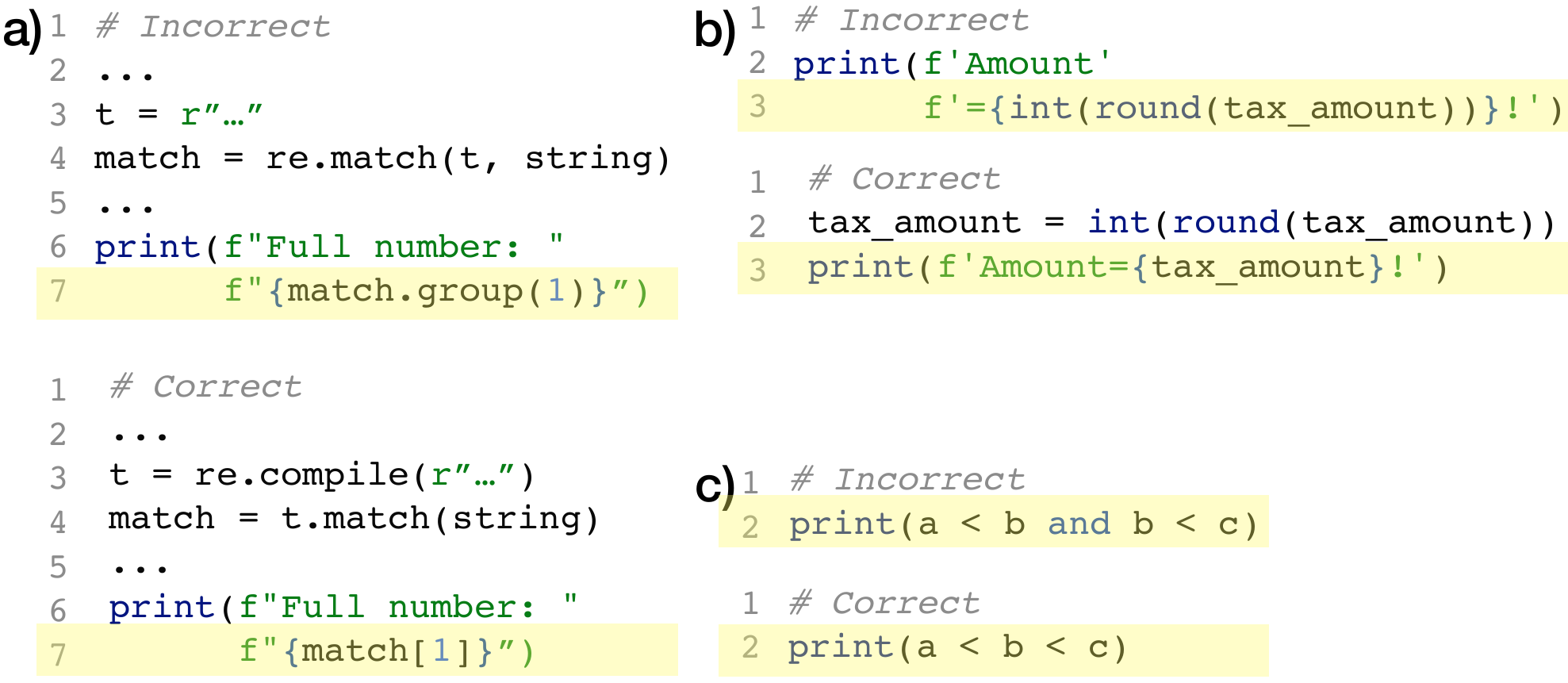}
    \vspace{-0.5cm}
    \caption{Examples of RQ2 Python cases: a, b) Too complex formatted string violation; c) Chained comparison. }
    \label{fig:rq2:cases:python}
    \vspace{-0.2cm}
\end{figure}

\subsubsection{\textbf{Too complex formatted string} (Top-1 Python unfixed issue)}
This issue appears if the student uses an \texttt{f-string} and calls something inside it. For instance, the example in \Cref{fig:rq2:cases:python}a calls the \texttt{group} function, and the example in \Cref{fig:rq2:cases:python}b calls the \texttt{int} and \texttt{round} functions. 
This can complicate the code and lead to potential bugs.

\textit{Summary.}
Almost all students write the correct solution at the beginning, but then, in an attempt to shorten it, they transfer some of their code into the \texttt{f-string}, which causes the issue.
At this stage, students might not yet be familiar with the concept of refactoring and therefore do it incorrectly.

\subsubsection{\textbf{Chained comparison} (Top-3 Python fixed issue)}
This issue appears if students use two comparisons instead of a single triple one (see \Cref{fig:rq2:cases:python}c).
Python allows chaining comparison operators as a helpful shorthand, but such a ``syntactic sugar'' is not common in other programming languages.

\textit{Summary.}
There are not many guidelines about code quality on the JetBrains Academy platform at the moment.
However, \texttt{Chained comparison} issue is covered by the theory in Task\#5920~\cite{ task5920}. 
This task provides several illustrative examples and clear explanations, which can be the reason why 83\% of the students fix it in their final attempts.

 \textbf{Discovered reasons to fix code quality issues}.  \textbf{(1)} When students exchange their solutions, they can both acquire new knowledge and borrow other students' issues; \textbf{(2)} If an issue is simple to correct, students mostly try to do it, even with issues that are not covered in the theory part; \textbf{(3)} However, if the issue is too difficult to understand, students may ignore it. In this case, theory can help.

\section{Anomalously Long Submission Series}\label{sec:long:series}

\begin{figure*}[t]
    \centering
    \includegraphics[width=\linewidth]{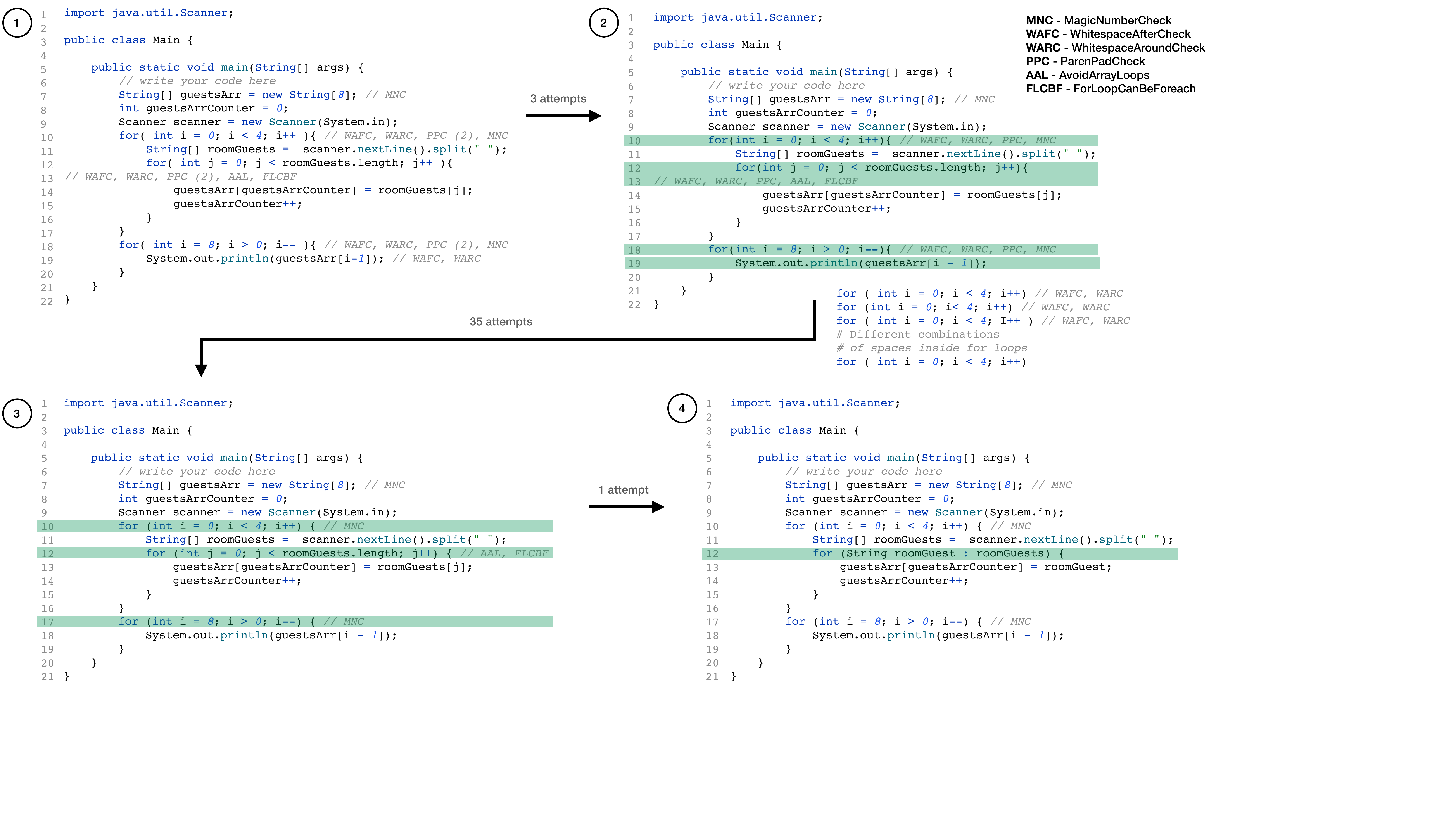}
    \vspace{-0.3cm}
    \caption{The full submission series for solving Task\#9261~\cite{task9261}, consisting of 39 attempts. The green color indicates places with the fixed code quality issues between attempts. Comments contain the list of code quality issues in the code line and their number in the brackets (if it contains more than one such issue). (1) Contains the first successful attempt, (4) --- the last attempt. Between (2) and (3) there were 35 attempts, the changes are shown next to the corresponding arrow, changing line 10.}
    \label{fig:rq3:long:series:java}
    \vspace{-0.6cm}
\end{figure*}

\begin{figure*}[t]
    \centering
    \includegraphics[width=\linewidth]{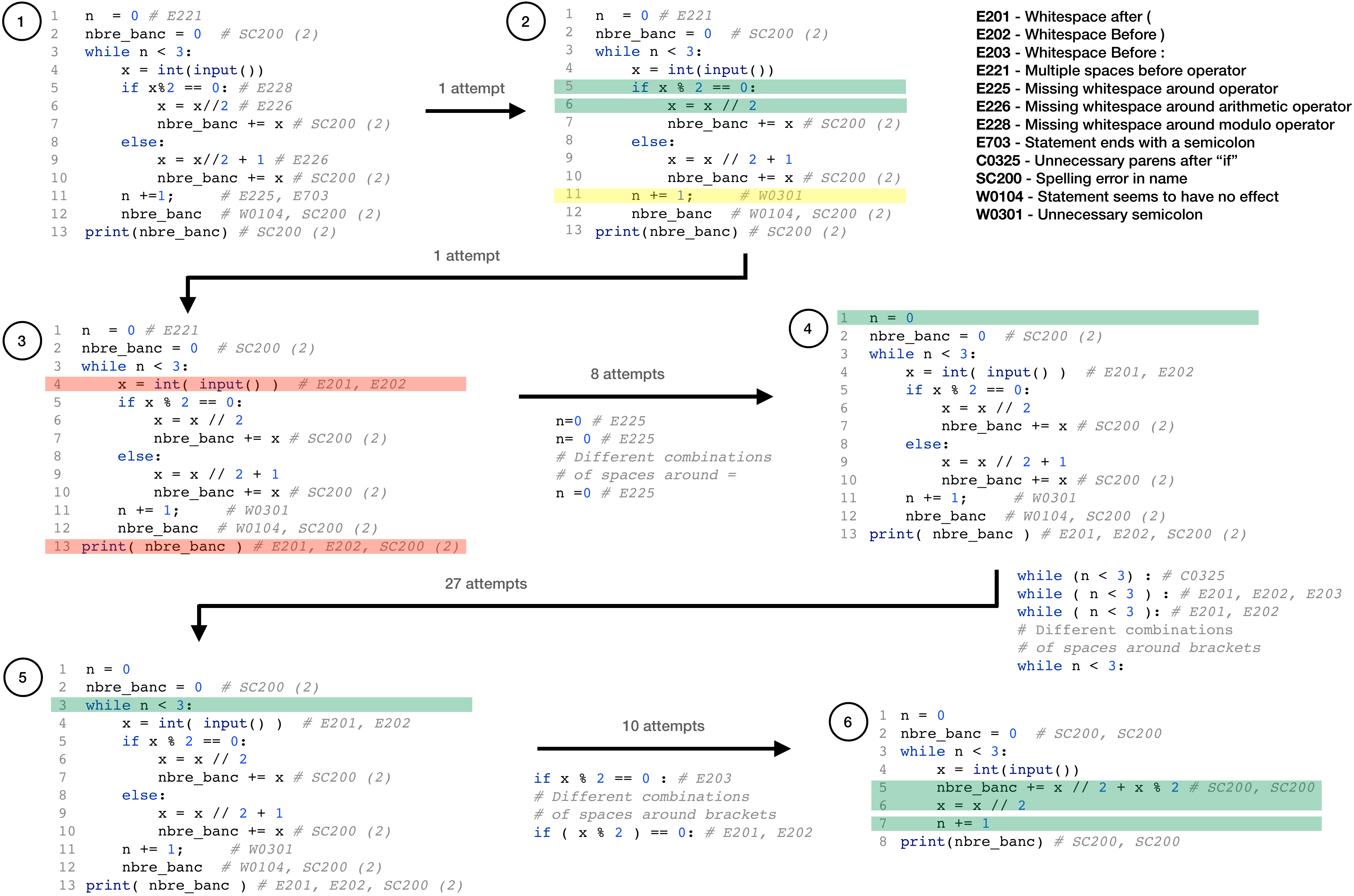}
    \vspace{-0.6cm}
    \caption{The full submission series for solving Task\#6462, consisting of 47 attempts. The green color indicates places with the fixed code quality issues between attempts. The yellow color indicates places where the initial issues were corrected, but new ones appeared. The red color indicates places where new code quality issues appeared. Comments contain the list of code quality issues in the code line and their number in the brackets (if it contains more than one such issue). (1) contains the first successful attempt, (6) --- the last attempt. Between (3) and (4) there were 8 attempts, the changes are shown next to the corresponding arrow, changing line 1. Between (4) and (5) there were 27 attempts, the changes are shown next to the corresponding arrow, changing line 3. Between (5) and (6) there were 10 attempts, the changes are shown next to the corresponding arrow, changing line 5.}
    \label{fig:rq3:long:series:python}
    \vspace{-0.6cm}
\end{figure*}

In the main body of this study, answering two research questions above, we did not take into account long submission series (longer than 5 successful submissions), which constitute about 5\% of the entire dataset. 
In this additional study, we manually analyze these long sequences to examine how students fix issues and why they might make numerous attempts. 
In the previous sections, we omitted code quality issues related to formatting, because they are less interesting and they would dominate the results in RQ1. 
However, in this part, we kept all formatting issues because they play a major role in these anomalously long sequences.

\subsection{Data}
For this analysis, we extracted all student submission series of length greater than 5 from the dataset described in~\Cref{sec:dataset}.
Then, we filtered out submissions made by internal JetBrains Academy bots, which resend all submissions for a specific task after some content or any of the tests change. 
From the standpoint of out dataset, these bots look like separate students that have suspiciously long artificial submission series of more than 1,000 attempts, which were sent one by one with less than a minute gap, but had quite different code inside (solutions of different students), making them easy to detect and remove.
Thus, the final data for this analysis consisted of only 5,180 submission series in Java and 2,786 in Python, with 5 to 50 attempts in each.

Finally, we calculated the frequency of different lengths of the remaining series.
For Java, most students have 6-10 attempts (about 92.6\%). 
However, several students solve tasks in 31, 39, and even 41 attempts.
For Python, the situation is similar --- about 93.9\% of the students have 6--10 attempts.
However, the maximum number of attempts for Python is slightly higher --- one of the students tried to pass the solution 47 times. 
At the same time, the rest of the students made no more than 29 attempts, which is lower than in Java.

\subsection{Java Case Study}
We chose the more interesting of the longest series to present as a case study.
This submission series has 39 attempts for Task\#9261~\cite{task9261}, where it is necessary to read several lines from the keyboard and output them in reverse order. 
The full history of correct submissions is presented in~\Cref{fig:rq3:long:series:java}.

The first successful attempt that passes all tests \circled{1} has 6 unique code quality issues and 19 issues in total. 
As we can see, almost all of them are about code formatting in \texttt{for} loops. 
To begin with, the student got rid of indentation issues step by step within 3 attempts \circled{2}: \textbf{(1)} firstly, deleted extra space between the open bracket and \texttt{int} in lines 10 and 18; \textbf{(2)} then added missing spaces in the \texttt{guestsArr} array index in line 19; \textbf{(3)} finally, removed extra spaces between \texttt{i++}, \texttt{j++}, \texttt{i--}, and the closing bracket in lines 10, 12, and 18.
We assume that the student knew exactly how to fix the issue at each attempt, as it was fixed in several places via one change.

However, after these changes, the \texttt{for} loops still contains formatting issues: \textit{Whitespace After} and \textit{Whitespace Around}. 
In fact, the loops do not contain a space after the \texttt{for} keyword and before the opening curly bracket.
Probably, because the JetBrains Academy interface highlights only the line with the issue, the student did not understand their exact positions and went through various options for 35 attempts.
Finally, the student found the right combination \circled{3}.
In the end, a more serious issue was fixed --- the \texttt{for} loop was replaced with \texttt{forEach} \circled{4}.
The final code looks more readable, but still contains a few issues --- all of them do not affect the final grade, so the student finished this task with the highest score.

\subsection{Python Case Study}

For Python, we analyzed the longest submission series with 47 attempts.
The student tried to solve Task\#6462~\cite{task6462}, which requires reading three integer numbers from the user input (the number of students in three classes) and calculating the minimum number of desks to be purchased if at most two students may sit at any desk.
The full history of the correct submissions is presented in~\Cref{fig:rq3:long:series:python}.

The first successful attempt that passes all the tests \circled{1} has 7 unique code quality issues and 17 issues in total, and almost all of them are about code formatting in arithmetic expressions. 
Thus, the student tried to correct them first and succeeded (lines 5, 6 and 11 in \circled{2}).
However, another issue in line 11 (\texttt{E703}) was unexpectedly substituted with (\texttt{W0301}).
Both of these issues point to an extra semicolon at the end of line 11, but they were detected by different linters inside the Hyperstyle tool~\cite{birillo2021hyperstyle} and deduplicated in a wrong way.

Probably, this platform behavior confused the student, so in the third attempt \circled{3}, the student added extra spaces around \texttt{input} (line 4) and inside \texttt{print} (line 13), which produced four new formatting issues.
Then, the student switched to the first line and tried to correct wrong spaces in 8 attempts \circled{4}.
It is interesting to note that the student tried different combinations of spaces, while there were already similar lines in the code that did not contain this issue (\textit{e.g.}, lines 2, 4, 6, etc.).
Besides the highlighting of the exact issue position, a clearer message from the tool could help with this task faster, since the current version of the hint is too general: \textit{missing whitespace around operator}.

Next, the student added parentheses around the \texttt{while} statement and tried to add spaces in the correct way. 
After 27 attempts, the extra brackets were removed and the solution between \circled{4} and \circled{5} did not change. 
Finally, we see the same problem, but with the \texttt{if} statement between \circled{5} and \circled{6}. 
For both of these cases, the student tried to correct the issues that they themselves added, experimenting with parentheses and spaces.
Perhaps, the problem is that students can not see their submissions history and roll back code with code quality issues they made to the previous state.

As a result, after all corrections, the student's final solution \circled{6} is much cleaner and shorter than in \circled{1}. 
However, it should be noted that issues related to incorrect spelling and those that do not affect the final score have not been corrected.

 \textbf{Summary}. \textbf{(1)} Issue hints must be carefully adapted in code quality assessment tools because students can spend a lot of attempts to fix an issue just by going through all the possible options; \textbf{(2)} Students can change correct code to incorrect code. Perhaps, in this case, it is necessary to additionally indicate these cases to the student and highlight them among other issues; \textbf{(3)} Some students find it difficult to correct formatting issues (spaces, brackets, etc) in the Web editor if the exact position is not set, resulting in a lot of attempts; \textbf{(4)} Students are reluctant to correct issues that do not affect the final code quality grade.

\section{Implications}

Let us reiterate the main practical results from the analysis of cases in both RQs and anomalously long submission series.

\textbf{Theoretical part.} The theoretical part of a task should be complete and covering different cases. First and foremost, it should not contain quality issues in itself, and, equally important, it should cover all the specific sub-cases that will appear in the practical part.

\textbf{Extensive testing.} Test cases for practical tasks should be exhaustive and account for different possible solutions. Perhaps, this is not feasible to predict from the start, so this part should be revisited after some time by analyzing the most popular solutions.

\textbf{Pre-written templates}. If the task contains a pre-written template, the template also should not contain code quality issues itself. 
Moreover, the template should not be too restrictive and force a particular solution, but instead enable different solution strategies.

\textbf{Reporting issues.} The overall results of our study indicate that the reporting of code quality issues to students is paramount in the education process. 
In a lot of cases, these reports allow the students to fix the issues in their initial attempts, thus practically instilling code quality guidelines.

\textbf{Rewarding students.} Specifically, we found that the \textit{grade} for code quality (that does not directly influence the overall result of the course) is a good incentive for a student to put effort into the task, and, conversely, some issues that do not influence the grade, remain unfixed.

\textbf{Adapting the hints.} As mentioned before in Section~\ref{sec:background}, a problem with some code quality tools is that their results are unadapted for novices. In our study, we also found that students sometimes struggle with understanding what exactly the issue is. Overall, even when adopting messages from tools, developers should be very careful with wording. Also, for simple and understandable formatting issues, finding their exact place in the marked line can also present a challenge. 
\section{Threats to Validity}\label{sec:threats}

In this work, we conducted a large-scale study of code quality issues, however, it has some limitations.
In RQ1, the presence of each specific issue is not uniform among different tasks, which means that there are tasks where a specific issue is more prevalent, and these tasks can inflate its prevalence in general.
However, this is normal: different tasks target different constructs and concepts in the language, and are thus more prone to different types of issues.
For this reason, issues such as \textit{Missing break in switch} are more prevalent in certain tasks, while more general issues like \textit{Redundant import check} are more universal.
We believe that this does not invalidate the results of our study, since we used a diverse dataset of \totalsteps tasks that target a wide variety of topics.

Also, the results of this work are not generalizable to all MOOC platforms, since they were obtained on a specific platform --- JetBrains Academy.
We conducted the manual analysis to look into case studies for RQ1 and RQ2, as well as long series, however,  these findings are not directly generalizable, since they depend on specific tasks and environments.
At the same time, we received and shared a lot of insights that can be useful to the community in different settings.

\section{Conclusion and Future Work}\label{sec:conclusion}

In this paper, we carried out an analysis of the code quality of successful student submissions from the JetBrains Academy platform.
We studied the most popular code quality issues among the users' initial successful submissions, analyzed the dynamics of fixing these issues on a large scale, and analyzed the most interesting cases manually.

We found that students introduce various kinds of code quality issues, with the real root causes being not only the students themselves, but also external factors, such as an incorrect task or learning materials, as well as obscure code quality issues messages. The issues also significantly differ in terms of how often they get fixed. Some issues are fixed by the majority of students if the issue is simple to understand or easy to fix. However, some issues are fixed more rarely, and there even exist issues that are more common in the last attempts than in the first. This can happen due to students copying other students' solutions after solving the task, or when they try to shorten the code without the proper knowledge of refactoring practices. Finally, in about 5\% of the cases, the submission series can be more than 5 attempts long, and we found that these long series of attempts often consist of unsuccessfully fixing simple formatting issues.
The supplementary materials for the paper are available online~\cite{artifacts}.

We believe that this study will be useful to teachers, developers of code quality tools, as well as MOOC educational platforms in general. One can focus on specific issues highlighted in this work to make sure they are addressed, but it is also crucial to improve existing materials and tools, in particular so that they do not introduce issues themselves.
In future work, we plan to study in greater detail the influence of grading on code quality, as well as differences in code quality issues between the Web environment and the IDE.

\bibliographystyle{ieeetran}
\balance
\bibliography{paper}

\end{document}